\documentclass[%
superscriptaddress,
 reprint,
nofootinbib,
nobibnotes,
 amsmath,amssymb,
 aps,
]{revtex4-2}

\usepackage{tikz}
\usepackage{tkz-euclide}
\usepackage[compat=1.1.0]{tikz-feynman}

\usepackage{float}

\usepackage{multirow}
\usepackage{caption}
\usepackage{subcaption}
\usepackage{relsize}

\usepackage{graphicx}
\usepackage{dcolumn}
\usepackage{bm}
\usepackage{hyperref}
\usepackage{slashed}

\begin{document}


\title{mass of $1^{-+}$ four-quark--hybrid mixed states}
\author{Shuang-Hong Li}
\email{leesh@zju.edu.cn}
\author{Ze-Sheng Chen}%
\email{ventuschen@zju.edu.cn}
\author{Hong-Ying Jin}
\email{jinhongying@zju.edu.cn}
\affiliation{Zhejiang Institute of Modern Physics, Department of Physics, Zhejiang University, Hangzhou, 310027, China}%
\author{Wei Chen}%
\email{chenwei29@mail.sysu.edu.cn}
\affiliation{School of Physics, Sun Yat-Sen University, Guangzhou 510275, China}%

\date{\today}

\begin{abstract}
	We calculate the masses of $J^{PC}=1^{-+}$ light exotic mesons by QCD sum rules; the masses are extracted from four-quark--hybrid mixing correlation functions. We construct several $1^{-+}$ four-quark currents and hybrid currents, and get two masses around $1.2\text{-}1.4\text{GeV}$ and $1.45\text{-}1.67\text{GeV}$; they can be identified as $\pi_1(1400)$ and $\pi_1(1600)$. 
\end{abstract}

\maketitle

\section{\label{intro}Introduction}

The study of exotic meson states has a long history. Glueball, hybrid meson (quark-antiquark with excited gluon), and multi-quark were predicted after the establishment of quantum chromodynamics. Many heavy exotic hadrons have been confirmed experimentally in the last decade. There are also two light exotic mesons with $J^{PC}=1^{-+}$ have been found since the end of the last century. However, the structure of these two mesons is still unclear.

Lattice QCD and most phenomenological methods show that the mass of $1^{-+}$ hybrid is around $1.7\text{-}2.1\text{GeV}$ \cite{meson1,meson3,hybrid3,hybrid4}, and it prefers to decay into S- and P-wave mesons. However, the corresponding candidates $\pi_1(1400)$ and $\pi_1(1600)$\cite{pdg} are lighter than predictions. Besides, $\pi_1(1400)$ only decays into $\eta\pi$, while Ref~\cite{su3} indicates that in the limit of $SU(3)$ flavor symmetry, the $1^{-+}$ hybrid decay into $\eta\pi$ is forbidden. This result implies $\pi_1(1400)$ may be a multi-quark state.

Some authors consider the $1^{-+}$ exotic mesons as tetraquark (diquark-antidiquark structure) or four-quark molecule (mesons bound state). Ref~\cite{4quark3} exhaust all $1^{-+}$ tetraquark configurations and get two masses around $1.6\text{GeV}$ and $2.0\text{GeV}$ for two types of quark contents. Ref~\cite{mole1} gives $1^{-+}$ four-quark molecule mass around $1.4\text{-}1.5\text{GeV}$, but the result receives large uncertainty from the instanton density. Ref~\cite{hybrid3} re-examines the studies about $1^{-+}$ exotic mesons, and predicts $1.7\text{GeV}$ for tetraquark configuration and $1.3\text{GeV}$ for  molecule configuration.

However, if $\pi_1(1400)$ is a four-quark state \footnote{We call both tetraquark and molecule configurations as four-quark states when the distinction between them are not important in discussion.}, then there should be unobserved mesons with strangeness 2 and 2 units of charge\cite{su3}. Besides, there is a subtle problem for the study about four-quark state. The four-quark current, usually can be identified as two meson currents, may easily couple to two mesons. It then causes difficulty to extract the correct information about the resonance when the two-mesons states give large contributions to the correlator. Ref~\cite{4quark_2} indicates that the four-quark diagrams with no singularity at $s=(\sum_{i=1}^4 m_i)^2$ are not relevant to the four-quark state but relevant to two free mesons ($m_i$ is the quark mass). But the validity of this criterion is not clear up to now.

Since the gluon can couple to quark-antiquark pair, the four-quark states may easily mix with the hybrid state, as long as the symmetry permits, especially for light four-quark and hybrid, since $g\gtrsim1$ in low energy scale. Ref~\cite{hybrid3} conjecture that the mixing of four-quark molecule with tetraquark and/or hybrid can explain the strange decay model of $\pi_1(1400)$. The mixing scenario may also explain two $1^{-+}$ states with close masses and the absence of four-quark states with strangeness 2 and 2 units of charge.

In this paper, we will try to evaluate the mass of $1^{-+}$ mesons from four-quark--hybrid correlation functions. The off-diagonal correlator certainly is weaker than the diagonal correlators. However, since the four-quark and hybrid have different decay models, the background contribution of two free mesons is strongly suppressed, which relatively enhances the contribution of the resonance that couples to both currents. This approach is not sensitive to the mixing strength, but our goal is to extract the state with the most prominent resonance signal in the correlators. We will give further discussion at the end of this paper. We think the mass evaluation based on four-quark--hybrid correlation function is convincible, and we get the masses of $1^{-+}$ states consistent with experiments.

\section{QCD sum rules for four-quark--hybrid}
\subsection{Currents and Renormalization}
The QCD sum rules~\cite{sum0,*sum00,sum02,sum04} is an effective way to research hadron properties. For mass evaluation, the main task is to calculate the correlation function of two currents. By decay models of $\pi_1(1400)$ and $\pi_1(1600)$~\cite{pdg}, we choose $\eta\pi$, $\eta'\pi$, $\rho\pi$, and $b_1\pi$ as four-quark current configurations. The currents for $\eta\pi$ and $\eta'\pi$ in the $SU(3)$ flavor limit are:
\begin{equation}
	\begin{split}
		J_{\eta^{(\prime)}\pi}^\mu =& (\bar{u}\gamma^5 \gamma^\mu u + \bar{d} \gamma^5 \gamma^ \mu d +\theta \bar{s} \gamma^5 \gamma^ \mu s)(\bar{u}\gamma^5 u - \bar{d}\gamma^5 d),\\	
		J_{\eta^{(\prime)}\pi}^{\mu \nu} =& (\bar{u}\gamma^5 \gamma^\mu u + \bar{d} \gamma^5 \gamma^ \mu d +\theta \bar{s} \gamma^5 \gamma^ \mu s)(\bar{u} \gamma^5 \gamma^ \nu u - \bar{d} \gamma^5 \gamma^ \nu d)\\
		&\ + \{\mu \leftrightarrow \nu\},
	\end{split}
	\label{eq1}
\end{equation}
with $\theta=-2$ for $\eta\pi$ and $1$ for $\eta'\pi$. The currents couple to $1^{-+}\ b_1\pi$ are:
\begin{equation}
	\begin{split}
		J_{b_1\pi}^{\mu} &= \epsilon^{\mu \alpha \beta \eta} \ (\bar{u} \sigma_{\alpha \beta} d \ \bar{d} \gamma^5 \gamma_\eta u - \bar{d} \sigma_{\alpha \beta} u \ \bar{u} \gamma^5 \gamma_\eta d), \\
		J_{b_1\pi}^{\mu\nu} &= i\epsilon^{\mu \nu \alpha \beta} \ (\bar{u} \sigma_{\alpha \beta} d \ \bar{d} \gamma^5 u - \bar{d} \sigma_{\alpha \beta} u \ \bar{u} \gamma^5 d).
	\end{split}
	\label{eq3}
\end{equation}
And the current couples to $\rho\pi$ \cite{mole1} is:
\begin{equation}
	J_{\rho\pi}^{\mu \nu} = i \epsilon^{\mu \nu \alpha \beta} \ (\bar{u} \gamma_\alpha d \ \bar{d} \gamma^5 \gamma_\beta u - \bar{d} \gamma_\alpha u \ \bar{u} \gamma^5\gamma_\beta d).
	\label{rp}
\end{equation}
Here $\epsilon^{0123} = +1$, $\sigma^{\mu\nu} = \frac{i}{2} [\gamma^\mu,\gamma^\nu]$.
The hybrid currents are:
\begin{equation}
	\begin{split}
		J_H^\mu &= i  (\bar{u}\ G^{\mu \nu} \gamma_\nu u - \bar{d}\ G^{\mu \nu} \gamma_\nu d), \\
		J_H^{\mu\nu} &= \bar{u} (G^{\mu\alpha} {\sigma_\alpha}^\nu - G^{\nu\alpha} {\sigma_\alpha}^\mu) u -\{u \leftrightarrow d \}.
	\end{split}
	\label{eq4}
\end{equation}

Here we always write $gT^a G^{a\mu\nu}$ as $G^{\mu\nu}$ for simplicity.

\begin{figure}
	\centering
	\includegraphics[width=0.48\textwidth]{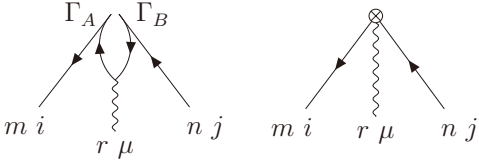}
	\caption{Renormalization of four-quark current at $O(g)$. The 4-quark vertex is split a little to make it clear how quark lines are connected. Left (right) diagram corresponding to Eq.~(\ref{eq6}) (Eq.~(\ref{eq7})).}
	\label{fig1}
\end{figure}

The four-quark currents need to be renormalized\cite{book_q,book_r}. The general four-quark current can be written as:
\begin{equation}
	J_{4q} = \overline{\Psi}_a  \Gamma_A  \Psi_b \ \overline{\Psi}_c \Gamma_B  \Psi_d.
	\label{eq5}
\end{equation}

Here $a, b, c, d$ are flavor indices; $\Gamma_A$, $\Gamma_B$ are general $\gamma$ and color matrices. The $1/\epsilon$-pole at $O(g)$ exist when $b=c$ and/or $a=d$. Assuming $b=c$ and $a\neq d$, we consider the zero-momentum insertion Green function:
\begin{equation}
	\langle0| J_{4q} \ \Psi_{a,i}^m \ A^{r \mu} \ \overline{\Psi}_{d,j}^n|0 \rangle.
	\label{eq6}
\end{equation}

Here, $i$, $j$ are spin indices; $m$, $r$, $n$ are color indices. The $ 1/\epsilon$-pole is canceled by (see Fig.~\ref{fig1}):
\begin{equation}
	\langle0| (J_1+J_2)\ \Psi_{a,i}^m \ A^{r \mu} \ \overline{\Psi}_{d,j}^n|0 \rangle,
	\label{eq7}
\end{equation}
with:
\begin{equation}
	\begin{split}
		J_1 &= -\frac{1}{\epsilon}\frac{m}{32\pi^2}  \overline{\Psi}_a \Gamma_A G^{\alpha\beta}  \sigma_{\alpha\beta} \Gamma_B  \Psi_d,\\
		J_2 &= \frac{1}{\epsilon}\frac{1}{48\pi^2} \overline{\Psi}_a \Gamma_A D_\alpha G^{\alpha\beta} \gamma_\beta \Gamma_B \Psi_d.
	\end{split}
	\label{eq8}
\end{equation}

Here, m is the mass of the quark in the loop, $D_\alpha^{ab} = \partial_\alpha \delta^{ab} + g f^{acb} A_\alpha^c$ is covariant derivative at adjoint representation. Eq.~(\ref{eq8}) is evaluated at dimension  $D=4-2\epsilon$. As the $J_1$ vanishes at massless limit, we use it only for evaluating $m \langle \bar{q}q\rangle$ contributions in this paper.

The renormalized four-quark current at $O(g)$ then can be written as:
\begin{equation}
	(J_{4q})_R = J_{4q} + J_1 +J_2.
	\label{eq9}
\end{equation}
By Eq.~(\ref{eq8}) and (\ref{eq9}), the renormalized currents for $\eta^{(\prime)}\pi$, $b_1\pi$ and $\rho\pi$ can be easily obtained.


\subsection{four-quark--hybrid correlation functions}
By operator product expansion~\cite{sum0,*sum00,sum02,sum04} (OPE), the correlation function can be expressed as a sort of condensates with Wilson coefficients. Before evaluating the correlation functions, we first discuss the dispersion relation\cite{sum0,*sum00,sum02,sum04}, which is slightly modified when two currents are different. Consider the correlation function of currents $J_a$ and $J_b$; for simplicity, assuming the involved states have same quantum number $J^{PC}$, suppressing the Lorentz indices, we have:
\begin{equation}
	\begin{split}
		\mathit{\Pi}(q)&=i\int d^4 x\ e^{iqx}\langle0|T\{J_{a}(x)J_b^{\dagger}(0)\}|0\rangle,\\
		&=\int_{0}^{\infty}ds\ \rho(s)\frac{\mathcal{P}(q)}{s-q^2-i \epsilon},\\
		&=\mathcal{P}(q)\bigg[\text{PP}\int_0^\infty ds\,\frac{\rho(s)}{s-q^2} + i \pi \rho(q^2)\bigg].
	\end{split}
	\label{eq10}
\end{equation}
Here $\text{PP}$ means principal part. We write:
\begin{equation}
	\langle0|J_a(0)|n\rangle\langle n|J_b^\dagger(0)|0\rangle=\mathcal{P}(q)f_a(q^2)f_b^*(q^2).
\end{equation}

Here $|n\rangle$ are on-shell states with momentum $q$; $\mathcal{P}(q)$ corresponds to the tensor structure of $\mathit{\Pi}(q)$; $f_a(q^2)$ and $f_b(q^2)$ are coupling constants of the two currents to $|n\rangle$ respectively. Define a factor:
\begin{equation}
	P=(-1)^{(N+M)}.
	\label{eq11}	
\end{equation}

Here $N=0,1,2$ is the number of anti-hermitian currents; $M=0$ ($1$) if $\mathit{\Pi}(q)$ is even (odd) under exchange $q\leftrightarrow -q$. Then we can write:
\begin{subequations}
	\begin{equation}
		\rho(s)=\sum_n \delta(s - m_n^2) \text{Re} [f_a(s)f_b^*(s)],
		\label{eq12a}
	\end{equation}
	for $P=1$, and:
	\begin{equation}
		\rho(s)= \sum_n\delta(s - m_n^2) i\, \text{Im} [f_a(s)f_b^*(s)],
		\label{eq12b}
	\end{equation}
	\label{eq12}
\end{subequations}
for $P=-1$.

When $J_a=J_b$, the $\rho(s)$ reduces to the familiar form $|f_a(s)|^2$ by Eq.~(\ref{eq12a}). We give the derivation of Eq.~(\ref{eq10})-(\ref{eq12}) in Appendix\ref{appA}, which is based on Ref~\cite{sum04}. When $J_a\neq J_b$, $\rho(s)$ or $\text{Im}\rho(s)$ may not be positive, and may even change the sign when $s$ variant.

For currents given by Eq.~(\ref{eq1})-(\ref{eq4}), to extract the $1^{-+}$ vector state contribution, we can generally write:
\begin{equation}
	\begin{split}
		\langle0|J_a^\mu(0)|V\rangle &=  \epsilon^\mu f_a(q^2),\\
		\langle0|J_b^{\mu\nu}(0)|V\rangle &= (q^\mu \epsilon^\nu \pm q^\nu \epsilon^\mu) f_b(q^2).
	\end{split}
	\label{eq13}
\end{equation}

Here, $|V\rangle$ is a vector state with momentum $q$ and polarization $\epsilon^\mu$, the $\pm$ refers to symmetry or anti-symmetry tensor currents. Then we have:
\begin{equation}
	\begin{split}
		\mathit{\Pi}_{a,b}^{\mu\nu\rho}(q) &= i\int d^4 x e^{iqx}\langle0|T\{J_{a}^{\mu\nu}(x)J_b^{\dagger\rho}(0)\}|0\rangle, \\
		&= (q^\mu \mathcal{P}^{\nu\rho} \pm q^\nu \mathcal{P}^{\mu\rho})\mathit{\Pi}_{a,b}^{T,V}(q^2) + \ldots\ \ .
	\end{split}
	\label{eq14}
\end{equation}

Here, $\mathcal{P}^{\mu\nu}=g^{\mu\nu} - \frac{q^\mu q^\nu}{q^2}$, the ellipsis refers to terms irrelevant with vector states. The superscript $T$ and $V$ in $\mathit{\Pi}_{a,b}^{T,V}(q^2)$ indicate that $J_a$ is a tensor current and $J_b$ is a vector current. In the massless limit, for currents given by Eq.~(\ref{eq1})-(\ref{eq4}), only the correlation functions with one vector current and one tensor current have non-vanish perturbative diagrams. We only consider this type of correlation function. By Eq.~(\ref{eq10}) and (\ref{eq12}), isolate the lowest resonance pole, we have:
\begin{equation}
	\frac{1}{\pi}\text{Im}\mathit{\Pi}_{a,b}(s)\simeq \delta(s-m^2) \text{Re}(f_a f_b^*) + \theta(s-s_0)\rho(s).
	\label{eq15}
\end{equation}

Here, $s_0$ is the continuum threshold, $P=1$ in this case.

\begin{figure}
	\centering
	\includegraphics[width=0.48\textwidth]{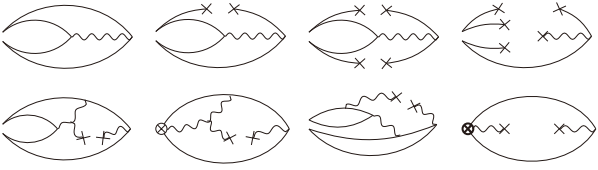}
	\caption{Typical diagrams. The 4-quark vertices are split a little to make it clear how quark lines are connected.}
	\label{fig2}
\end{figure}

For $\langle GG\rangle$ term, the two-loop renormalization of the four-quark currents should be considered, which is far beyond the scope of this paper. Since our purpose is only to evaluate the mass, we only need to extract the corresponding counterterms to cancel the non-local pole $\text{log}/\epsilon$ (corresponding to the last diagram in Fig.~\ref{fig2}). This diagram originates from the current:
\begin{equation}
	\overline{\Psi} (\overleftarrow{\nabla}G + G \overrightarrow{\nabla})\Psi= \partial(\overline{\Psi}G \Psi) + i \overline{\Psi} DG  \Psi,
\end{equation}
which is the counterterm that emerges in the renormalization at two-loop level (at $O(g^3)$). Here, the $\gamma$-matrices and Lorentz indices are suppressed, $\nabla^\mu$ is covariant derivative. The term $\partial (\overline{\Psi}G \Psi)$ gives a non-zero contribution to the correlation function. For $J_{\eta^{(\prime)}\pi}^\mu$ and $J_{b_1\pi}^\mu$, the corresponding counterterms are not unique (see Table.~\ref{tb3}), but different choices did not cause any visible difference in mass prediction.

For dimension-10 condensate, some diagrams need the identity (in massless limit):
\begin{subequations}
	\begin{equation}
		(\nabla^\mu \nabla^\nu \Psi_i)^a =\frac{ig\ \Gamma_{ij}^{\mu\nu\alpha\beta}}{2(d+2)} G_{\alpha\beta}^n(T^n)^{ab}\Psi_j^b,
	\end{equation}
with:
	\begin{align}
		\Gamma^{\mu\nu\alpha\beta}=&-(d^2-4d+8)g^{\mu\alpha}g^{\nu\beta}+\gamma^{\mu\nu\alpha\beta}+g^{\mu\nu}\gamma^{\alpha\beta}\nonumber \\
		&+(d-4)\gamma^{\mu\alpha}g^{\nu\beta}-(d-2)\gamma^{\nu\alpha}g^{\mu\beta}. 
	\end{align}
	\label{8expand}
\end{subequations}
Here, $\gamma^{\mu\nu\alpha\beta}=\gamma^{[\mu}\gamma^\nu\gamma^\alpha\gamma^{\beta]}$.

Since the $\text{Im}\mathit{\Pi}_{a,b}(q^2)$ in Eq.~(\ref{eq14}) is not positive definite, we define the spectrum density by requiring the perturbative part is positive, i.e., $\rho(s)>0$ when $s\rightarrow\infty$. We give the OPE results, counterterms, and all diagrams in Appendix\ref{appB}. The OPE calculations are performed up to dimension-10 condensates; we write a Mathematica package~\cite{package} to deal with the tedious calculation. The results are obtained at $\overline{\text{MS}}$ scheme, and the $\gamma^5$ is treated by BMHV scheme~\cite{BMHV1,feyncalc}.


\subsection{Numerical analysis}
To evaluate the mass of the lowest resonance, it commonly uses the Borel (Laplace) transfomation\cite{gauss,sum01,sum02}. By Eq.~\ref{eq15}, we have:
\begin{equation}
	\begin{split}
		\frac{1}{\pi}\int_{0}^{\infty} ds\ e^{-s\tau}\ \text{Im}\mathit{\Pi}_{a,b}(s)\simeq&\ \text{Re}(f_a f_b^*) e^{-m^2 \tau}\\
		&+\int_{s_0}^{\infty} ds\ e^{-s\tau}\rho(s).
	\end{split}
\end{equation}

And one can get the mass from the ratio of moments:
\begin{equation}
	\mathcal{R}_n = \frac{\mathcal{M}_{n+1}(\tau,s_0)}{\mathcal{M}_n(\tau,s_0)}\simeq m^2,
	\label{eq17}
\end{equation}
with:
\begin{equation}
	\mathcal{M}_n(\tau,s_0)=\int_0^{s_0} ds\ s^n\ e^{-s\tau}\ \text{Im}\mathit{\Pi}_{a,b}(s).
\end{equation}

Here, two free parameters $\tau$ and $s_0$ are involved. The typical $s_0$ is around the mass square of the next resonance. The range of $\tau$ is determined by requiring $\mathcal{M}_0(\tau,s_0)/\mathcal{M}_0(\tau,\infty)>0.7$, and the contribution of the highest dimensional condensate is less than $10\%$ to satisfy single-pole dominance and convergence of OPE. This gives the Borel window\cite{sum02} $\tau_1<\tau<\tau_2$. The stability criteria~\cite{miniqcd} is also used to constraint the value of $s_0$ and  $\tau$.

\begin{figure*}
	\centering
	\includegraphics[width=\textwidth]{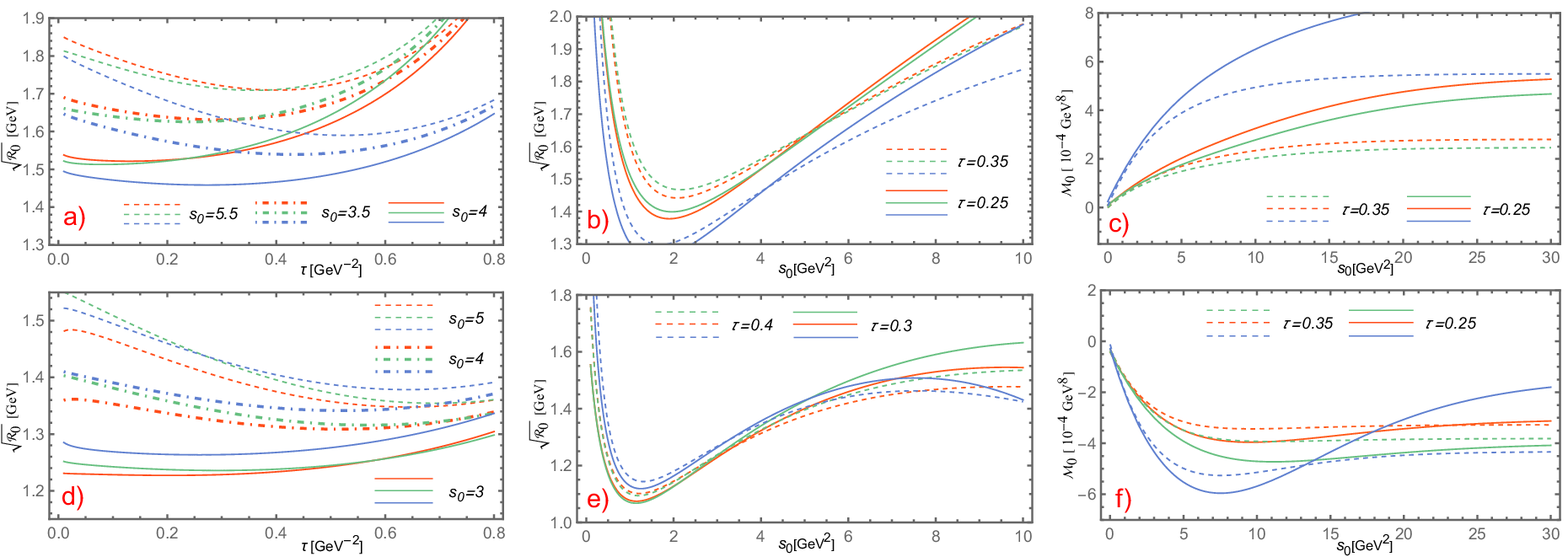}
	\caption{\label{fig3}Mass predictions and moments for different correlation functions. Here we set $k_8=k_{10}=3.5$. The colors represent the four-quark configurations: red for $\eta\pi$, green for $\eta'\pi$, and blue for $b_1\pi$. (a): Mass versus $\tau$ for $\mathit{\Pi}_{\eta^{(\prime)}\pi,H}^{V,\ T}(q^2)$ and $\mathit{\Pi}_{b_1\pi,H}^{T,\ V}(q^2)$; (b): mass versus $s_0$ for $\mathit{\Pi}_{\eta^{(\prime)}\pi,H}^{V,\ T}(q^2)$ and $\mathit{\Pi}_{b_1\pi,H}^{T,\ V}(q^2)$; (c): $\mathcal{M}_0(\tau,s_0)$ versus $s_0$ for $\mathit{\Pi}_{\eta^{(\prime)}\pi,H}^{V,\ T}(q^2)$ and $\mathit{\Pi}_{b_1\pi,H}^{T,\ V}(q^2)$; (d): mass versus $\tau$ for $\mathit{\Pi}_{\eta^{(\prime)}\pi,H}^{T,\ V}(q^2)$ and $\mathit{\Pi}_{b_1\pi,H}^{V,\ T}(q^2)$; (e): Mass versus $s_0$ for $\mathit{\Pi}_{\eta^{(\prime)}\pi,H}^{T,\ V}(q^2)$ and $\mathit{\Pi}_{b_1\pi,H}^{V,\ T}(q^2)$; (f): $\mathcal{M}_0(\tau,s_0)$ versus $s_0$ for $\mathit{\Pi}_{\eta^{(\prime)}\pi,H}^{T,\ V}(q^2)$ and $\mathit{\Pi}_{b_1\pi,H}^{V,\ T}(q^2)$; $\rho(s)$ change the sign at s around $7\text{-}10\text{GeV}^2$.}
\end{figure*}

To evaluate the mass, we use the value of condensates by Ref.~\cite{miniqcd}:

\begin{equation}
	\begin{split}
		(m_u + m_d)\langle \bar{u}u+\bar{d}d\rangle=&- f_\pi^2 m_\pi^2;\ \ f_\pi=130\text{MeV},\\
		\langle \bar{u}u\rangle=\langle \bar{d}d\rangle=&-(0.276\text{GeV})^3,\\
		\langle GG\rangle=&0.07\text{GeV}^4,\\
		\langle \bar{q}Gq\rangle =&0.8 \text{GeV}^2 \langle \bar{q}q\rangle;\ \ q=u,d.\\
	\end{split}
\end{equation}

And $\Lambda=0.353 \text{GeV}$ by Ref.~\cite{hybrid3}. For dimension-6, -8, and -10 condensates, the factorization deviation factors must be included:
\begin{equation}
	\begin{gathered}
		\langle \bar{q}q\rangle^2\rightarrow \ k_6\langle \bar{q}q\rangle^2, \\
		\langle \bar{q}q\rangle \langle\bar{q}Gq\rangle\rightarrow\ k_8 \langle \bar{q}q\rangle \langle\bar{q}Gq\rangle,\\
		\langle\bar{q}Gq\rangle^2 \rightarrow\ k_{10}  \langle\bar{q}Gq\rangle^2.
	\end{gathered}
\end{equation}

Here, $k_6\approx3$ by Ref.~\cite{violation6}\footnote{The $\rho\alpha_s\langle \bar{\psi}\psi\rangle^2$ in Ref.~\cite{miniqcd} already contain the deviation factor. But we find it is less stable to fix $\alpha_s$ first, so we will not use this value.}, the $k_8$ and $k_{10}$ are not clear; Ref.~\cite{factorization} indicate they are around 2-5. We first choose $k_8=k_{10}=3.5$ to fix Borel window and $s_0$, then consider two extremes $k_8=k_{10} =2$ and $k_8=k_{10} =5$ to get conservative results. We show the results for $k_8=k_{10}=3.5$ in Fig.~\ref{fig3}, and give the figures for $k_8=k_{10} =2$ and $k_8=k_{10} =5$ in Appendix\ref{appB}. The conservative ranges of masses are given in Table.~\ref{tb1}. Since the $s$-quark can not couple to the hybrid currents given by Eq.~(\ref{eq4}) directly, the contribution of $s$-quark is small. Our results are insufficient to distinguish $\eta\pi$ and $\eta'\pi$.

To obtain the mass, for specific $s_0$, if the Borel window exists, we choose the average value in the Borel window. If the Borel window does not exist, we determine the mass by stability criteria. We then get the range of mass spanned by $s_0$.

For $\mathit{\Pi}_{\eta^{(\prime)}\pi,H}^{V,\ T}(q^2)$ and $\mathit{\Pi}_{b_1\pi,H}^{T,\ V}(q^2)$, the Borel window and $s_0$ stability  do not exist. The $\tau$-stability is achieved when $s_0\approx5.5$ for $\mathit{\Pi}_{\eta^{(\prime)}\pi,H}^{V,\ T}(q^2)$, and when $s_0\approx4.5$ for $\mathit{\Pi}_{b_1\pi,H}^{T,\ V}(q^2)$ (the curve for different $\tau$ intersect at these $s_0$). We choose $4\leq s_0\leq 6$ for these configurations. The $s_0$ stability at $s_0\approx2$ in Fig.~\ref{fig3}(b) and $s_0\approx1$ in Fig.~\ref{fig3}(e) are not physical, because these $s_0$ are lower than the corresponding mass squared.

When $s_0=4$, the $\mathit{\Pi}_{\eta^{(\prime)}\pi,H}^{T,\ V}(q^2)$ has Borel window around  $0.33\text{-}0.36$, while the $\mathit{\Pi}_{b_1\pi,H}^{V,\ T}(q^2)$ has $\tau_1\approx\tau_2\approx0.38$. We treat the latter one as no Borel window. The $\tau$ stability is achieved when $s_0\approx3$ (Fig.~\ref{fig3}(e)), but the $\tau$ stability becomes worse when $s_0\approx5$ (Fig.~\ref{fig3}(d)). We choose $3\leq s_0\leq 5$ for these configurations.

The $\mathit{\Pi}_{\rho\pi,H}^{T,\ V}(q^2)$ has no dimension-8 and -10  contributions at leading order, and is dominated by dimension-6 condensate. The result is then sensitive to factorization deviation; it is hard to get a convincible value. It is interesting to mention that the $1^{-+}$ hybrid decays into $f_1 \pi$ and $b_1 \pi$, but not to $\rho\pi$ by flux-tube model\cite{hybrid5}. These facts imply the $1^{-+}$ hybrid may prefer to couple to $b_1\pi$ than $\rho\pi$.

The main uncertainties in Table.~\ref{tb1} are from the factorization deviation factors, which cause a small overlap between $\pi_1(1400)$ and $\pi_1(1600)$ interpretation. For the same deviation factors, there is a roughly $0.2\text{GeV}$ gap between them. So we conclude there are two $1^{-+}$ states.

\begin{table}
	\caption{Mass predictions by each correlation function in the unit of GeV. The lower  bound is obtained by using the lower bound of $s_0$ and set $k_8=k_{10}=2$; the upper bound is obtained by using the upper bound of $s_0$ and set $k_8=k_{10}=5$.}
	\label{tb1}
	\begin{subtable}{\linewidth}
		\caption{\label{tb1a}Results can be interpret as $\pi_1(1600)$.}
		\begin{ruledtabular}
			\renewcommand{\arraystretch}{1.5}
			\begin{tabular}{ccc}
				$\displaystyle \mathit{\Pi}_{\eta\pi,H}^{V,\ T}(q^2)$&$\displaystyle \mathit{\Pi}_{\eta'\pi,H}^{\ V,\ T}(q^2)$&$\displaystyle \mathit{\Pi}_{b_1\pi,H}^{T,\ V}(q^2)$\\ \hline
				$\displaystyle 1.45\text{-}1.77$&$\displaystyle 1.45\text{-}1.77$&$\displaystyle 1.36\text{-}1.67$
			\end{tabular}
		\end{ruledtabular}
	\end{subtable}
	\begin{subtable}{\linewidth}
		\caption{\label{tb1b}Results can be interpret as $\pi_1(1400)$.}
		\begin{ruledtabular}
			\renewcommand{\arraystretch}{1.5}			
			\begin{tabular}{ccc}
				$\displaystyle \mathit{\Pi}_{\eta\pi,H}^{T,\ V}(q^2)$&$\displaystyle \mathit{\Pi}_{\eta'\pi,H}^{\ T,\ V}(q^2)$&$\displaystyle \mathit{\Pi}_{b_1\pi,H}^{V,\ T}(q^2)$\\ \hline
				$\displaystyle 1.18\text{-}1.41$&$\displaystyle 1.19\text{-}1.43$&$\displaystyle 1.17\text{-}1.46$
			\end{tabular}
		\end{ruledtabular}
	\end{subtable}
\end{table}

Recall that we fix the sign of $\rho(s)$ by requiring the perturbative contribution is positive. However, for $\mathit{\Pi}_{\eta^{(\prime)}\pi,H}^{T,\ V}(q^2)$ and $\mathit{\Pi}_{b_1\pi,H}^{V,\ T}(q^2)$, the $\rho(s)$ is negative for $s\lesssim8$. A closer look finds that for $\tau$ in the Borel window, the moment $\mathcal{M}_0(\tau,s_0)$ is always negative. That means the contribution of the resonance is negative. It should not be surprised since $\text{Re}[f_a f^*_b]$ for different intermediate states may have different signs.

In Table.~\ref{tb1}, the mass prediction from $\mathit{\Pi}_{b_1\pi,H}^{V,\ T}(q^2)$ is close to $\pi_1(1400)$, while the $b_1\pi$ is absent in the decay of $\pi_1(1400)$. Since $b_1\pi$ is just on the mass threshold of $\pi_1(1400)$, $b_1\pi$
channel may be forbidden or suppressed by little kinematic phase space in $\pi_1(1400)$'s decay.

It should note that in Eq.~(\ref{eq15}), we assume only one resonance pole dominates, but there have two close $1^{-+}$ states: $\pi_1(1400)$ and $\pi_1(1600)$. However, different currents may prefer to couple to a certain one, and it is unlikely that all currents in Eq.~(\ref{eq1})-(\ref{eq4}) couple to these two states with the same preference. By results in Table.~\ref{tb1}, we think each case is dominated by one state. 


\section{Discussion and Conclusion}

As we mentioned in the Introduction, we think the mass evaluation based on the four-quark--hybrid correlation is convincible.  Since whatever $|J(x)\rangle$ is, it can always be written as a sort of physical states. We can write:
\begin{equation}
	\begin{split}
		J(x)\sim \sum A(x)&+ \sum\int d^4q_1 d^4q_2\, e^{-i(q_1+q_2)x}\\
		&\times \psi(q_1,q_2)\, \phi_1(q_1)\phi_2(q_2)+\dots\ \  .
	\end{split}
	\label{eq25}
\end{equation}

Here, the $A(x)$, $\phi_1(q_1)$, and $\phi_2(q_2)$ are effective fields of mesons; the $\psi(q_1,q_2)$ is a certain wave function of two free mesons in momentum space; the ellipsis refers to three and more particle states. In principle, the free mesons terms should not be excluded. The correlation function then can be expressed as:

\begin{equation}
	\begin{split}
		\int d^4x\,& e^{iqx} \langle T\{J_a(x)J_b(0)\}\rangle\sim \\
		&\begin{tikzpicture}[baseline=-\the\dimexpr\fontdimen22\textfont2\relax]
			\begin{feynman}[inline=(a0)]
				\vertex (a0);
				\vertex [right=1.2cm of a0](a1);
				\diagram*[small]{
					(a1)--[thick](a0),
				};
			\end{feynman}
		\end{tikzpicture}+
		\begin{tikzpicture}[baseline=-\the\dimexpr\fontdimen22\textfont2\relax]
			\begin{feynman}[inline=(a0)]
				\vertex (a0);
				\vertex [right=0.8cm of a0](a1);
				\vertex [right=0.6cm of a1](a2);
				\diagram*[small]{
					(a1)--[bend left=40]a0[empty dot]--[bend left=40](a1),
					(a1)--[thick](a2),
				};
			\end{feynman}
		\end{tikzpicture}+\begin{tikzpicture}[baseline=-\the\dimexpr\fontdimen22\textfont2\relax]
			\begin{feynman}[inline=(a0)]
				\vertex (a0);
				\vertex [left=0.8cm of a0](a1);
				\vertex [left=0.6cm of a1](a2);
				\diagram*[small]{
					(a1)--[bend right=40]a0[empty dot]--[bend right=40](a1),
					(a1)--[thick](a2),
				};
			\end{feynman}
		\end{tikzpicture}+\begin{tikzpicture}[baseline=-\the\dimexpr\fontdimen22\textfont2\relax]
			\begin{feynman}[inline=(a0)]
				\vertex (a0);
				\vertex [right=1.2cm of a0](a2);
				\diagram*[small]{
					a0[empty dot]--[bend left=40]a2[empty dot]--[bend left=40]a0,
				};
			\end{feynman}
		\end{tikzpicture}\\
	&+\dots \ \ .
	\end{split}
\label{eq26}
\end{equation}

Here, sum over the different particles for each term is understood; the empty dots refer to two free mesons.

Consider the energy scale around the mass of $1^{-+}$ mesons, for $J_a=J_b=\text{four-quark current}$, since it can be identified as two meson currents, the last term in Eq.~\ref{eq26} may contribute considerably. For $J_a=J_b=\text{hybrid current}$, one may safely ignore the multi mesons terms in Eq.~\ref{eq25} at first approximation. There are many studies about the $1^{-+}$ hybrid~\cite{hybrid3,hybrid4,hybrid7} based on the first current of Eq.~\ref{eq4}. However, if there are two exotic mesons with close masses, focusing on one current is not enough.

To compromise, one can choose $J_a=$ four-quark current and $J_b=$ hybrid current. As long as the second term in Eq.~\ref{eq26} is small, it is reasonable only to consider the first term. Since different currents may prefer to couple to different exotic mesons, and it is easy to construct many four-quark currents, it is possible to extract a certain exotic meson from these correlators. By comparing the results with experiments, one can check the validity of this argument.

The results in Table.~\ref{tb1} and Fig.~\ref{fig3} show that the two $1^{-+}$ states can be identified as $\pi_1(1400)$ and $\pi_1(1600)$. Since the results are obtained from four-quark--hybrid correlators, these states may be mixed states of four-quark and hybrid. The mixing strength can not obtain from the off-diagonal correlators, which needs the information of the diagonal correlators. However, it may be contaminated by two free mesons.

By Eq.~(\ref{eq5})-(\ref{eq9}), the mixing of four-quark and hybrid may be quite common. For the correlators in this paper, the contribution of perturbative diagrams is quite small, which implies the mixing is highly nonperturbative. The mixing scenario of $1^{-+}$ four-quark and hybrid may also explain the absence of four-quark states with strangeness 2 and 2 units of charge: the pure light four-quark state is unstable, it must mix with the hybrid to get stability. If it is true, the iso-scalar four-quark state should mix with the hybrid or glueball to get stability. Our calculations is an attempt to test this scenario.

In conclusion, the Table.~\ref{tb1} gives mass predictions with $1.2\text{-}1.4\text{GeV}$ for $\pi_1(1400)$ and $1.45\text{-}1.67\text{GeV}$ for $\pi_1(1600)$, which agree with PDG\cite{pdg} ($1.35\text{GeV}$ for $\pi_1(1400)$ and $1.66\text{GeV}$ for $\pi_1(1600)$). But it should note that a recent coupled channel analysis of COMPASS data concluded that there is only one $1^{-+}$ state with mass $1.56\text{GeV}$\cite{1-+pole}. So there are large uncertainties that need to be cleared both in theory and experiments; this topic is beyond this paper.

\appendix

\onecolumngrid

\section{\label{appA}Dispersion relation in generalized situation}

Consider the correlation function of two currents (suppress the Lorentz indices):
\begin{equation}
	\begin{split}
		\mathit{\Pi}(q)&=i\int d^4x e^{iqx}\langle0|T\{J_a(x)J_b^\dagger(0)\}|0\rangle,\\
		&=i\int d^4x e^{iqx}\big[\theta(x^0)\langle 0| J_a(x)J_b^\dagger(0)|0\rangle+\theta(-x^0)\langle 0| J_b^\dagger(0)J_a(x)|0\rangle\big].
	\end{split}
	\label{aeq1}
\end{equation}

Inserting a complete set of on-shell states:
\begin{equation}
	\sum_n\int \frac{d^4p}{(2\pi)^3}\theta(p^0)\delta(p^2-m_n^2)|n\rangle\langle n|=1.
\end{equation}

We obtain:
\begin{equation}
	\mathit{\Pi}(q)= \sum_n\int d^4x \, e^{iqx}\int \frac{d^4p}{(2\pi)^3}\mathcal{P}_n(p)\Big(i\theta(x^0)\theta(p^0)\delta(p^2-m_n^2)e^{-ipx}A(p^2) +i\theta(-x^0)\theta(p^0)\delta(p^2-m_n^2)e^{ipx}A^\dagger(p^2)\Big).
	\label{aeq3}
\end{equation}
Here we write $\langle0|J_a(0)|n\rangle\langle n|J_b^\dagger(0)|0\rangle=\mathcal{P}_n(p)A(p^2)$, $p$ is the momentum of state $|n\rangle$, $\mathcal{P}_n(p)$ is only relevant with the tensor structure of $\mathit{\Pi}(q)$.

Separating real and imaginary part, and changing the integral variable, we write the terms in big parenthesis  as:
\begin{equation}
	\begin{split}
		\big[&i\theta(x^0)\theta(p^0)\delta(p^2-m_n^2)+(-1)^{N+M}i\theta(-x^0)\theta(-p^0)\delta(p^2-m_n^2)\big]e^{-ipx}\, \text{Re}[A(p^2)] \\
		&+\big[i\theta(x^0)\theta(p^0)\delta(p^2-m_n^2)-(-1)^{N+M}i\theta(-x^0)\theta(-p^0)\delta(p^2-m_n^2)\big]e^{-ipx}\, i\text{Im}[A(p^2)].
	\end{split}
\end{equation}

Here $N=0,1,2$ is the number of antihermitian currents; $\mathcal{P}_n(-p)=(-1)^M \mathcal{P}_n(p)$, or equivalently $\mathit{\Pi}(-q)=(-1)^M \mathit{\Pi}(q)$. Eq.~(\ref{aeq3}) then can be written as $\mathit{\Pi}(q)=\iint \mathcal{P}_n(p)(\text{Re}+\text{Im})$ briefly. Under replacement $x\rightarrow-x$, $p\rightarrow-p$, and $q\rightarrow-q$, it becomes:
\begin{equation}
	(-1)^M\mathit{\Pi}(q)=(-1)^{N+2M}\iint \mathcal{P}_n(p)\text{Re} \, -\,(-1)^{N+2M}\iint \mathcal{P}_n(p)\text{Im}.
\end{equation}

Thus for $N+M=\text{even}$, the $\text{Im}[A(p^2)]$ term vanishes, while for $N+M=\text{odd}$, the $\text{Re}[A(p^2)]$ term vanishes.  By:
\begin{equation}
	\int \frac{d^4p}{(2\pi)^3} \Big(i\theta(x^0)\theta(p^0)\delta(p^2-m_n^2)+i\theta(-x^0)\theta(-p^0)\delta(p^2-m_n^2)\Big)e^{-ipx}=\int \frac{d^4p}{(2\pi)^4}\frac{e^{-ipx}}{m_n^2-p^2-i\epsilon},
\end{equation}
we then can write:
\begin{equation}
	\begin{split}
		\mathit{\Pi}(q)&=\sum_n\frac{\mathcal{P}_n(q)}{m_n^2-q^2-i\epsilon}\Big(\frac{1+(-1)^{N+M}}{2}\text{Re}[A(q^2)]+\frac{1-(-1)^{N+M}}{2}\, i\text{Im}[A(q^2)]\Big),\\
		&=\int_0^\infty ds \, \sum_n \delta(s-m_n^2)\frac{\mathcal{P}_n(q)}{m_n^2-q^2-i\epsilon}\Big(\frac{1+(-1)^{N+M}}{2}\text{Re}[A(q^2)]+\frac{1-(-1)^{N+M}}{2}\, i\text{Im}[A(q^2)]\Big).
	\end{split}
\end{equation}

So for $N+M=\text{even}$, the dispersion relation is:
\begin{equation}
	\frac{1}{\pi}\text{Im}\big[\mathit{\Pi}(q)\big]=\sum_n\delta(q^2-m_n^2)\text{Re}[A(q^2)],
\end{equation}
and for $N+M=\text{odd}$, the dispersion relation is:
\begin{equation}
	\frac{1}{\pi}\text{Re}\big[\mathit{\Pi}(q)\big]=-\sum_n\delta(q^2-m_n^2)\text{Im}[A(q^2)].
\end{equation}

\section{\label{appB}Additional Table and Figures}

\begin{table}[hb!]
	\caption{\label{tb2} The OPE Results. Here $m=m_u+m_d$, $\langle GG\rangle=g^2\langle G^{n\, \mu\nu}G_{\mu\nu}^n\rangle$, $\langle\bar{q}Gq\rangle=g\langle\bar{q}T^n G_{\mu\nu}^n\sigma^{\mu\nu}q\rangle$. The values in each column are the factors of corresponding terms, e.g., $\mathit{\Pi}_{\eta\pi,H}^{T,\ V}(q^2)=-1/12960\pi^5$ $ \alpha_s q^6 \text{log}\big(-q^2/\mu^2\big)+\ldots\ $. For $\mathit{\Pi}_{\eta^{(\prime)}\pi,H}^{\ V,\ T}(q^2)$ and $\mathit{\Pi}_{b_1\pi,H}^{V,\ T}(q^2)$, the relevant counterterms are not unique, the values in parenthesis corresponding to $\langle GG\rangle$ term are obtained by replacing $\displaystyle \partial^\alpha(\bar{q}G^{\mu\beta}\sigma_{\alpha\beta} q)$ by $\displaystyle \partial^\alpha(\bar{q}G_{\alpha\beta}\, \sigma^{\mu\beta} q)$ in Table.~\ref{tb3}.}
	\begin{ruledtabular}
		\renewcommand{\arraystretch}{1.6}
		\begin{tabular}{cccccccc}
			\\[-8pt]	
			&$\displaystyle \mathit{\Pi}_{\eta\pi,H}^{T,\ V}(q^2)$&$\displaystyle \mathit{\Pi}_{\eta'\pi,H}^{\ T,\ V}(q^2)$&$\displaystyle \mathit{\Pi}_{\eta\pi,H}^{V,\ T}(q^2)$&$\displaystyle \mathit{\Pi}_{\eta'\pi,H}^{\ V,\ T}(q^2)$&$\displaystyle \mathit{\Pi}_{b_1\pi,H}^{V,\ T}(q^2)$&$\displaystyle \mathit{\Pi}_{b_1\pi,H}^{T,\ V}(q^2)$&$\displaystyle \mathit{\Pi}_{\rho\pi,H}^{T,\ V}(q^2)$\\[5pt] \hline \\[-8pt]
			$\displaystyle \alpha_s q^6 \text{log}\Big(-\frac{q^2}{\mu^2}\Big)$	&$\displaystyle \frac{-1}{12960\pi^5}$&$\displaystyle \frac{-1}{12960\pi^5}$&$\displaystyle \frac{-1}{8640\pi^5}$&$\displaystyle \frac{-1}{8640\pi^5}$&$\displaystyle \frac{-1}{4320\pi^5}$&$\displaystyle \frac{-1}{8640\pi^5}$&$\displaystyle \frac{-1}{8640\pi^5}$\\[8pt]
			$\displaystyle \alpha_s q^2 \text{log}\Big(-\frac{q^2}{\mu^2}\Big)^2  m\langle\bar{q}q\rangle$	&$\displaystyle \frac{1}{72\pi^3}$&$\displaystyle \frac{1}{72\pi^3}$&$\displaystyle \frac{-1}{72\pi^3}$&$\displaystyle \frac{-1}{72\pi^3}$&$\displaystyle \frac{1}{36\pi^3}$&$\displaystyle \frac{-1}{36\pi^3}$&0\\[8pt]
			$\displaystyle \alpha_s q^2 \text{log}\Big(-\frac{q^2}{\mu^2}\Big)  m\langle\bar{q}q\rangle$&$\displaystyle \frac{-101}{432\pi^3}$&$\displaystyle \frac{-101}{432\pi^3}$&$\displaystyle \frac{23}{216\pi^3}$&$\displaystyle \frac{23}{216\pi^3}$&$\displaystyle \frac{-47}{108\pi^3}$&$\displaystyle \frac{29}{108\pi^3}$&$\displaystyle \frac{-1}{18\pi^3}$\\[8pt]
			$\displaystyle \alpha_s q^2 \text{log}\Big(-\frac{q^2}{\mu^2}\Big)^2  \langle GG\rangle$&$\displaystyle \frac{247}{995328\pi^5}$ &$\displaystyle \frac{-509}{995328\pi^5}$ &$\displaystyle \frac{281}{497664\pi^5}$ &$\displaystyle \frac{65}{497664\pi^5}$&$\displaystyle \frac{113}{248832\pi^5}$ &$\displaystyle \frac{223}{248832\pi^5}$&$\displaystyle \frac{181}{248832\pi^5}$\\[8pt]	
			$\displaystyle \alpha_s q^2 \text{log}\Big(-\frac{q^2}{\mu^2}\Big)  \langle GG\rangle$	&$\displaystyle \frac{-20147}{5971968\pi^5}$ & $\displaystyle \frac{14269}{5971968\pi^5}$&$\displaystyle \frac{-8861(-8765)}{1492992\pi^5}$ & $\displaystyle \frac{-3407(-2015)}{1492992\pi^5}$&$\displaystyle \frac{-3137(-2033)}{746496\pi^5}$&$\displaystyle \frac{-5899}{746496\pi^5}$&$\displaystyle \frac{-3547}{746496\pi^5}$\\[8pt]
			$\displaystyle \alpha_s \text{log}\Big(-\frac{q^2}{\mu^2}\Big) \langle\bar{q}q\rangle^2$&$\displaystyle \frac{56}{81\pi}$&$\displaystyle \frac{56}{81\pi}$&$\displaystyle \frac{-8}{27\pi}$&$\displaystyle \frac{-8}{27\pi}$&$\displaystyle \frac{32}{27\pi}$&$\displaystyle \frac{-20}{27\pi}$&$\displaystyle \frac{4}{27\pi}$\\[8pt]
			$\displaystyle \frac{1}{q^2} \langle \bar{q}q\rangle \langle\bar{q}Gq\rangle $	&$\displaystyle \frac{1}{18}$ &$\displaystyle \frac{1}{18}$ &$\displaystyle \frac{-1}{36}$ &$\displaystyle \frac{-1}{36}$&$\displaystyle \frac{1}{18}$ &$\displaystyle \frac{-1}{18}$&0\\[8pt]	
			$\displaystyle \frac{1}{q^4} \langle\bar{q}Gq\rangle^2 $&$\displaystyle \frac{49}{432}$&$\displaystyle \frac{49}{432}$&$\displaystyle \frac{-67}{648}$&$\displaystyle \frac{-67}{648}$&$\displaystyle \frac{55}{324}$&$\displaystyle \frac{-79}{432}$&0\\[8pt]
			$\displaystyle \frac{1}{q^4} \langle\bar{q}q\rangle^2\langle GG \rangle $&$\displaystyle \frac{1}{1944}$&$\displaystyle \frac{1}{1944}$&$\displaystyle \frac{-1}{7776}$&$\displaystyle \frac{-1}{7776}$&$\displaystyle \frac{1}{3888}$&$\displaystyle \frac{-1}{3888}$&0\\[8pt]
		\end{tabular}
	\end{ruledtabular}
\end{table}

\begin{table}[hb!]
	\caption{\label{tb3}Two-loop level counterterms (at $O(g^3)$). Here, e.g., $\displaystyle \frac{g^2}{\epsilon\,864\pi^4}\, \partial^\alpha(\bar{q}G^{\mu\beta}\sigma_{\alpha\beta} q)$ is counterterm for $J_{\eta\pi}^\mu$; $q=(u,d)^T$, and $\bar{q}=(\bar{u},-\bar{d})$. For $J_{\eta^{(\prime)}\pi}^\mu$ and $J_{b_1\pi}^\mu$, the counterterms are not unique. One can replace $\displaystyle \partial^\alpha(\bar{q}G^{\mu\beta}\sigma_{\alpha\beta} q)$ by $\displaystyle \partial^\alpha(\bar{q}G_{\alpha\beta}\, \sigma^{\mu\beta} q)$, which gives small change for $\langle GG\rangle$ terms in Table.~\ref{tb2}, but does not affect the mass prediction.}
	\begin{ruledtabular}
		\renewcommand{\arraystretch}{1.5}
		\begin{tabular}{cccccccc}
			\\[-10pt]	
			&$\displaystyle J_{\eta\pi}^{\mu \nu}$&$\displaystyle J_{\eta'\pi}^{\mu \nu}$&$\displaystyle J_{\eta\pi}^\mu$&$\displaystyle J_{\eta'\pi}^\mu$&$\displaystyle J_{b_1\pi}^{\mu}$&$\displaystyle J_{b_1\pi}^{\mu\nu}$&$\displaystyle J_{\rho\pi}^{\mu \nu}$\\[5pt] \hline \\[-8pt]
			$\displaystyle\partial^\mu(\bar{q}G^{\nu\alpha}\gamma_\alpha q)+\{\mu\leftrightarrow\nu\}$&$\displaystyle \frac{-7g^2}{\epsilon 3456\pi^4}$&$\displaystyle \frac{161g^2}{\epsilon 13824\pi^4}$&&&&&\\[8pt]
			$\displaystyle\partial^\mu(\bar{q}G^{\nu\alpha}\gamma_\alpha q)-\{\mu\leftrightarrow\nu\}$&&&&&&$\displaystyle \frac{17g^2}{\epsilon\,6912\pi^4}$&$\displaystyle \frac{-g^2}{\epsilon\,1728\pi^4}$\\[8pt]	
			$\displaystyle \partial^\alpha(\bar{q}G^{\mu\beta}\sigma_{\alpha\beta} q)$&&&$\displaystyle \frac{g^2}{\epsilon\,864\pi^4}$&$\displaystyle \frac{29g^2}{\epsilon\,1728\pi^4}$&$\displaystyle \frac{23g^2}{\epsilon\,864\pi^4}$&&\\[8pt]
		\end{tabular}
	\end{ruledtabular}
\end{table}

\begin{figure*}[ht!]
	\centering
	\includegraphics[width=\textwidth,height=230pt]{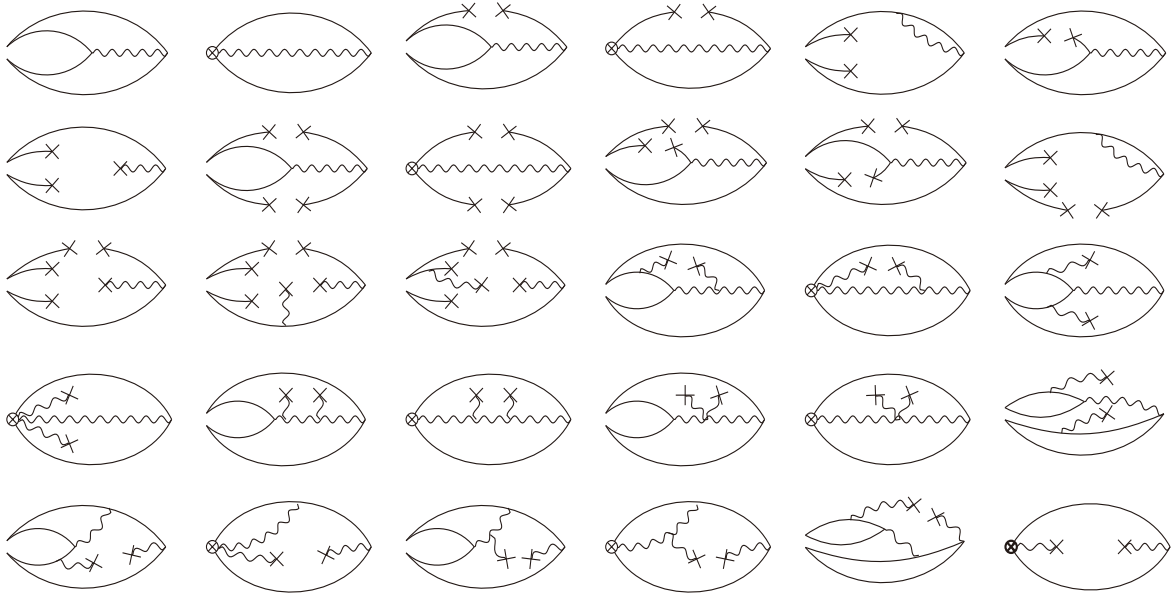}
	\caption{\label{Afig_4}Diagrams for four-quark--hybrid correlation functions, up to permutations of background gluons. For the first diagram in the second row, the identity $\langle\overline{\Psi}_i (x) G^{\alpha\beta}(0)\Psi_j(x)\rangle=-g^2\langle\overline{\Psi}\Psi\rangle^2/(36(D-1))\ (x^\mu\gamma^\nu-x^\nu\gamma^\mu)_{ji}$ is used. The diagrams in the last row are relevant with renormalization at two-loop level; the last diagram comes from the counterterms.}
\end{figure*}


\begin{figure*}[ht!]
	\centering
	\includegraphics[width=\textwidth]{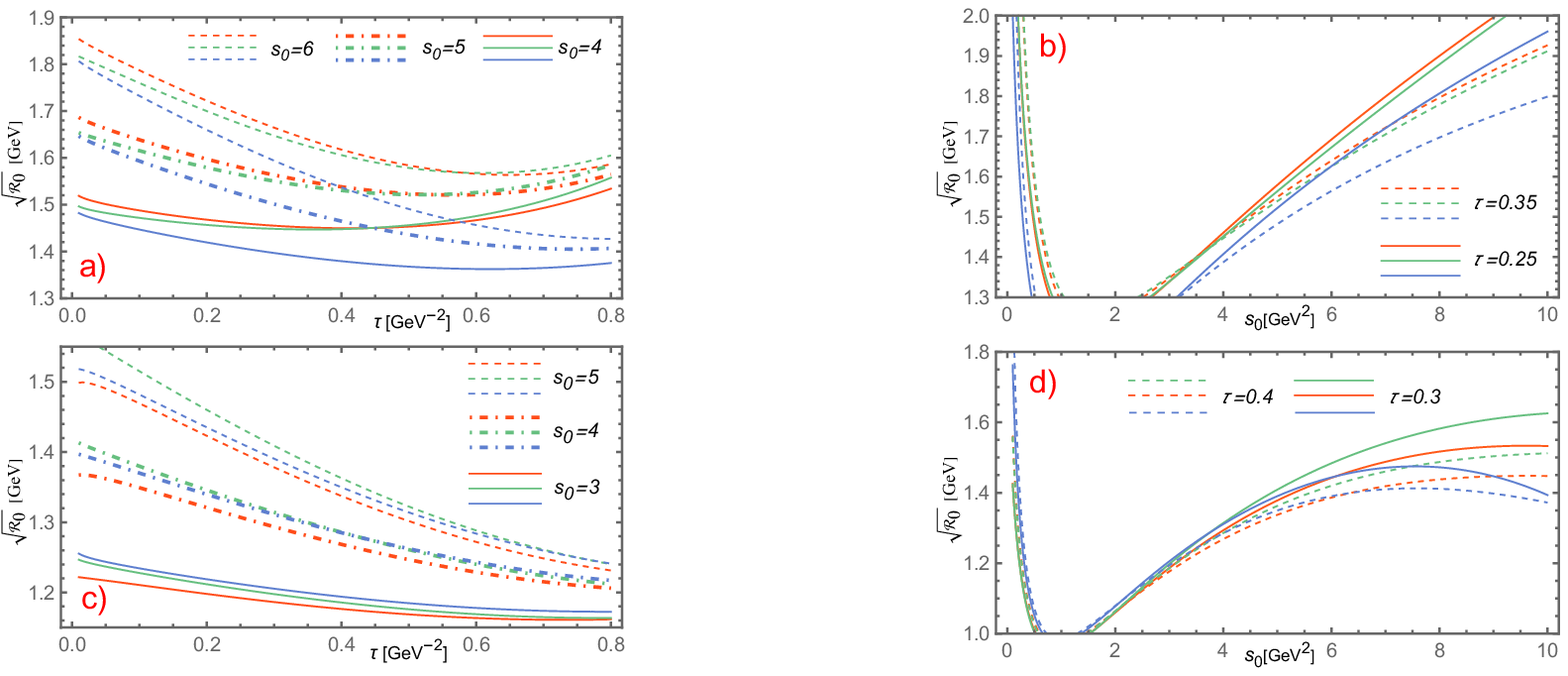}
	\caption{Mass predictions for different correlation functions. Here we set $k_8=k_{10}=2$. The colors represent the four-quark configurations: red for $\eta\pi$, green for $\eta'\pi$, and blue for $b_1\pi$. (a): Mass versus $\tau$ for $\mathit{\Pi}_{\eta^{(\prime)}\pi,H}^{V,\ T}(q^2)$ and $\mathit{\Pi}_{b_1\pi,H}^{T,\ V}(q^2)$; (b): mass versus $s_0$ for $\mathit{\Pi}_{\eta^{(\prime)}\pi,H}^{V,\ T}(q^2)$ and $\mathit{\Pi}_{b_1\pi,H}^{T,\ V}(q^2)$; (c): mass versus $\tau$ for $\mathit{\Pi}_{\eta^{(\prime)}\pi,H}^{T,\ V}(q^2)$ and $\mathit{\Pi}_{b_1\pi,H}^{V,\ T}(q^2)$; (d): Mass versus $s_0$ for $\mathit{\Pi}_{\eta^{(\prime)}\pi,H}^{T,\ V}(q^2)$ and $\mathit{\Pi}_{b_1\pi,H}^{V,\ T}(q^2)$.}
\end{figure*}


\begin{figure*}[ht!]
	\centering
	\includegraphics[width=\textwidth]{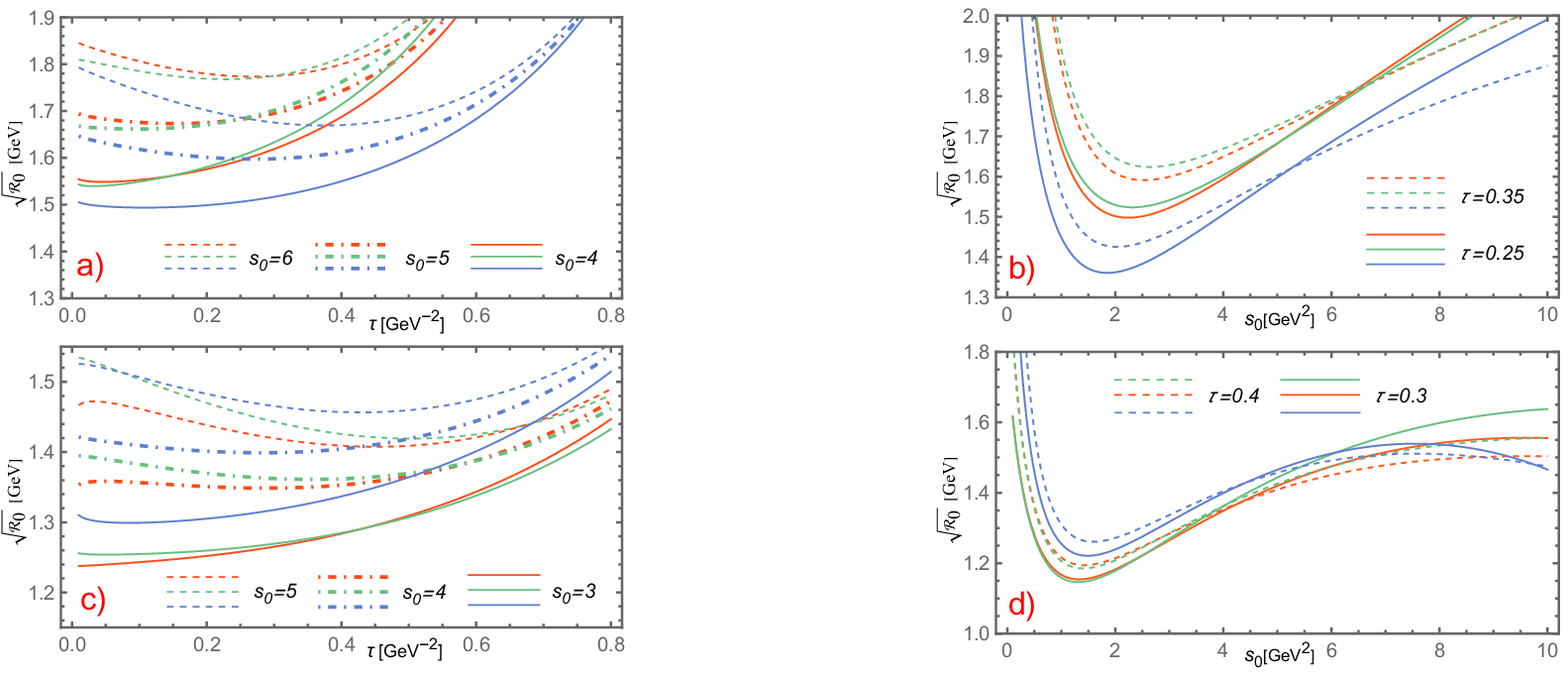}
	\caption{Mass predictions for different correlation functions. Here we set $k_8=k_{10}=5$. The colors represent the four-quark configurations: red for $\eta\pi$, green for $\eta'\pi$, and blue for $b_1\pi$. (a): Mass versus $\tau$ for $\mathit{\Pi}_{\eta^{(\prime)}\pi,H}^{V,\ T}(q^2)$ and $\mathit{\Pi}_{b_1\pi,H}^{T,\ V}(q^2)$; (b): mass versus $s_0$ for $\mathit{\Pi}_{\eta^{(\prime)}\pi,H}^{V,\ T}(q^2)$ and $\mathit{\Pi}_{b_1\pi,H}^{T,\ V}(q^2)$; (c): mass versus $\tau$ for $\mathit{\Pi}_{\eta^{(\prime)}\pi,H}^{T,\ V}(q^2)$ and $\mathit{\Pi}_{b_1\pi,H}^{V,\ T}(q^2)$; (d): Mass versus $s_0$ for $\mathit{\Pi}_{\eta^{(\prime)}\pi,H}^{T,\ V}(q^2)$ and $\mathit{\Pi}_{b_1\pi,H}^{V,\ T}(q^2)$.}
\end{figure*}

\vbox to .07\textheight{\vspace{1cm}}

\begin{equation*}
\end{equation*}

\twocolumngrid

\bibliography{Paper}

\end{document}